\documentclass[traditabstract,letter]{aa} % for the abstract without structuration 
\usepackage{graphicx}
\usepackage{txfonts}
\usepackage{natbib}
\begin{document}
\title{First firm spectral classification of an early-B \\ pre-main-sequence star: B275 in M17\thanks{Based on observations
    performed with the ESO {\it Very Large Telescope} on Cerro
    Paranal, Chile, as part of the X-shooter Science Verification
    program 60.A-9402(A).}}

\authorrunning{B.B. Ochsendorf et al.}
\titlerunning{First firm spectral classification of an early-B PMS star: B275 in M17}

%   \subtitle{I. Overviewing the $\kappa$-mechanism}

   \author{B.B. Ochsendorf\inst{1}
          \and
          L.E. Ellerbroek\inst{1}
          \and
          R. Chini\inst{2,3}
          \and
          O.E. Hartoog\inst{1}
          \and
          V. Hoffmeister\inst{2}
           \and
         L.B.F.M. Waters\inst{4,1}
           \and
         L. Kaper\inst{1}
          }

   \institute{Astronomical Institute Anton Pannekoek, University of
     Amsterdam, Science Park 904, P.O. Box 94249, 1090 GE Amsterdam,
     The Netherlands\\
              \email{ochsendorf@strw.leidenuniv.nl; L.Kaper@uva.nl}
         \and
             Astronomisches Institut, Ruhr-Universit\"{a}t Bochum,
             Universit\"{a}tsstrasse 150, 44780 Bochum, Germany
         \and
             Instituto de Astronom\'{i}a, Universidad Cat\'{o}lica del Norte, Antofagasta, 
             Chile
	 \and
	     SRON, Sorbonnelaan 2, 3584 CA Utrecht, The Netherlands}

   \date{Received; accepted}

% \abstract{}{}{}{}{} 
% 5 {} token are mandatory

   \abstract{The optical to near-infrared (300 -- 2500~nm)
     spectrum of the candidate massive young stellar object (YSO) B275,
     embedded in the star-forming region M17, has been observed with
     X-shooter on the ESO {\it Very Large Telescope}. The spectrum
     includes both photospheric absorption lines and emission features
     (H and Ca~{\sc ii} triplet emission lines, 1$^{\rm st}$ and
     2$^{\rm nd}$ overtone CO bandhead emission), as well as an
     infrared excess indicating the presence of a (flaring) circumstellar
     disk. The strongest emission lines are double-peaked with a peak
     separation ranging between 70 and 105~km~s$^{-1}$, and they provide information on the
     physical structure of the disk. The underlying photospheric
     spectrum is classified as B6--B7, which is significantly cooler than a
     previous estimate based on modeling of the spectral energy
     distribution. This discrepancy is solved by allowing for a larger
     stellar radius (i.e. a bloated star) and thus positioning the star above the main
     sequence. This constitutes the first firm spectral classification
     of an early-B pre-main-sequence (PMS) star. We discuss the position of B275 in the
     Hertzsprung-Russell diagram in terms of PMS
     evolution. Although the position is consistent with PMS tracks of heavily accreting protostars (\.M$_{\rm acc}
     \ga 10^{-5}$~M$_{\odot}$~yr$^{-1}$), the fact that the
     photosphere of the object is detectable suggests that the current
     mass-accretion rate is not very high.
     
   }

   \maketitle
%
%________________________________________________________________

\section{Introduction}

Observational and theoretical evidence is accumulating that the
formation process of massive stars is through disk accretion, similar
to low-mass stars. This persists despite the strong radiation pressure and
ionizing power produced by the massive young stellar object (YSO) that
may reverse the accretion flow and prevent matter from accreting onto
the forming star \citep[e.g.,][]{Keto06,Krumholz09}. Given the short
main-sequence lifetime of massive stars, the mass accretion rate must
be high \citep[up to $\sim 10^{-3}$~M$_{\odot}$~yr$^{-1}$,
][]{Hosokawa10} to ensure that the star is not leaving the main
sequence before the accretion process has finished.

Evidence of accretion must come from the
detection of circumstellar disks, and possibly bipolar jets, as
observed around forming low-mass stars \citep[e.g.,][]{Appenzeller89}. Disks
and outflows around massive YSO candidates are being reported
\citep[e.g.,][]{Chini04,Kraus10,Ellerbroek11},
but the physical properties of the forming massive stars remain
uncertain. The mass of the central object has to be estimated from the
emerging flux, and the direct detection of the photospheric spectrum
turns out to be very difficult at this early stage of evolution \citep[e.g.,][]{Testi10}.

% , and in accreting neutron stars and (supermassive) black holes \citep[e.g.,][]{Fender06}

%The latter authors use the definition that a massive YSO is a stellar object which is mid-IR bright and processes sufficient luminosity to form %a massive star ($>$ a few times $10^{3}$~L$_{\odot}$), but does not show evidence that an H~{\sc ii} region has begun to form (i.e., no %radio emission and no spatially extended mid-infrared emission). This is a rather broaddefinition; when adding some specific spectroscopic %signaturesindicating the presence of a disk and/or a hot photosphere, the sample of massive YSOs becomes much smaller. For only a few %massive YSO
Infrared surveys have revealed several hundred candidate massive YSOs, based on luminosity arguments \citep[e.g.,][]{Urquhart11}. 
A ($K$-band) spectrum has been obtained for only a few of these
\citep{Hanson97,Hanson02,Bik06}, and they show a red continuum,
likely due to hot dust, and an emission-line spectrum that includes
Br$\gamma$ and, often, CO 2.3~$\mu$m bandhead emission. The latter
emission can be modeled as being produced by a Keplerian rotating disk
surrounding the young, potentially massive star \citep{Bik04,Blum04,Wheelwright10}.

As massive stars show most spectral features in the UV and optical
ranges, the study of their photospheric properties would strongly
benefit from extending the spectral coverage as far to the blue as
possible. Obviously, extinction by the surrounding gas and dust makes
this an observational challenge. Only in rare cases have
spectra of candidate massive YSOs been obtained at optical wavelengths. \citet{Hanson97} 
obtained optical and near-infrared spectra of candidate massive 
YSOs in M17, one of the most massive nearby star-forming regions in
the Galaxy \citep{Hoffmeister08,Broos07,Povich09}. For the ``normal''
OB stars \citet{Hanson97} found a good correspondence between the
optical and $K$-band spectra, but the massive YSO optical spectra
remained inconclusive. For four massive YSO candidates, 
%(B226, B243, B268, B275) 
they registered the optical spectrum from 400 to 480~nm, indicating
a high mass and luminosity. The blue spectrum of the strong CO emission source B275
showed no definite photospheric features other than hydrogen, so that the nature of this source
remained uncertain. The spectral energy distribution (SED), though,
indicated spectral type late-O or early B, at an adopted distance of
1.3~kpc. 

  \begin{figure*}[!ht]
   \centering
%   \vbox{
  \includegraphics[width=\textwidth]{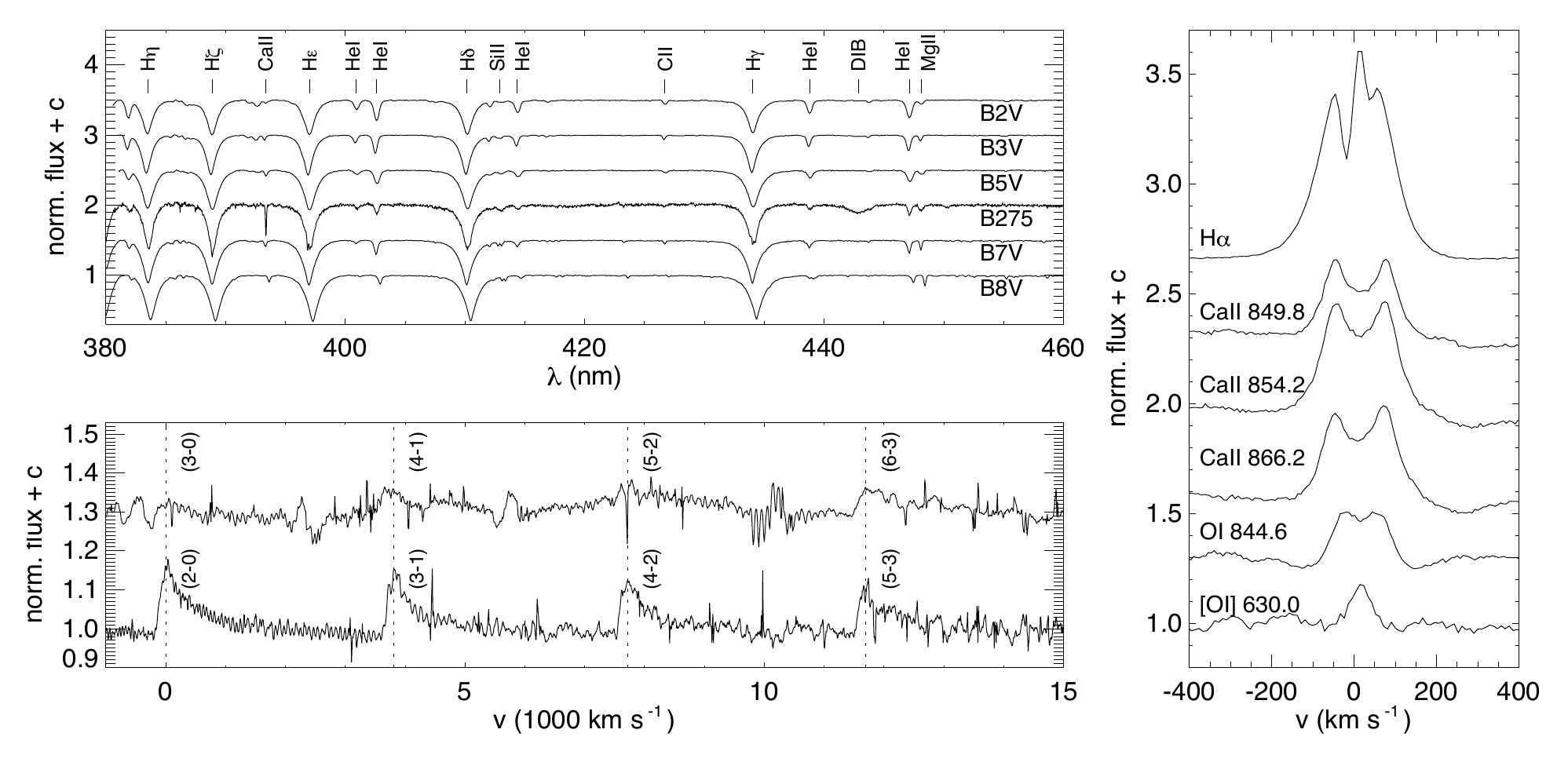}
%  \includegraphics[width=\textwidth]{spectralsequence_2.pdf}
 %  \includegraphics[width=\textwidth]{CaII.pdf}
 % obtained with the {\it Gray/Miller spectrograph} mounted on the 0.8m telescope of the {\it Dark Sky Observatory}
  % Line identifications are displayed above the spectra
% The He~{\sc i} / Mg~{\sc ii} line ratio is a sensitive diagnostic for spectral classification in this temperature range    and yields B6 (B7???) for B275.
      \caption{{\it Top left:} The blue spectrum of
    B275 in M17 shown next to B main-sequence-star 
    spectra \citep{Gray09}.  {\it Bottom left:} The 1$^{\rm st}$ and
    2$^{\rm nd}$ overtone CO emission bands. Zero velocity
    corresponds to the first component in the series (at 2294 and 1558~nm, 
    respectively). {\it Right:} A sample of the emission
    line profiles in the spectrum of B275. The Ca~{\sc ii} triplet
    lines and O~{\sc i} 845~nm are superposed on hydrogen Paschen series 
    absorption lines. The flux of the H$\alpha$ line is scaled down by a factor 5; the structure near
    the peak is a remnant of the nebular-line subtraction.}
              \label{fig:spectrum}
    \end{figure*}
% Note the relatively strong interstellar
%    Ca~{\sc ii} lines and the diffuse interstellar band at 443~nm in
  %  the spectrum of B275. 
We set out to exploit the high efficiency and broad wavelength
coverage of the new medium-resolution spectrograph X-shooter on the
ESO {\it Very Large Telescope} (VLT) to (i) detect the photospheric
spectrum of B275 in M17, (ii) determine its effective temperature
in order to place the candidate massive YSO unambiguously onto recent evolutionary
tracks, and (iii) search for ongoing accretion activity and
investigate the structure of the disk. 

%\begin{table}
%\begin{center}
%\begin{tabular}{|l|l|l|l|l|l|} \hline
%Arm & $\lambda$ range (nm) & $t_{\rm int}$ (s) & slit width ($''$) & S/N & $R$  \\ \hline%
%UVB & 300--550 & 4 $\times$ 685 & 1.6 & $\sim 74$ & 3300 %\\
%VIS & 550--1050 & 8 $\times$ 285 & 0.9 & $\sim 85$  & 8800 \\
%NIR & 1050--2500 & 12 $\times$ 220 & 0.9 & $\sim 60$  & 5600 \\ \hline
%\end{tabular}
%\caption{X-shooter instrument settings during the observations of B275
%  in M17 (RA(2000.0) = 18$^{\rm h}$20$^{\rm m}$25$\fs$13, DEC(2000.0) =
%  --16$^{\circ}$10$'$24$\farcs$56) on August 11, 2009 at 03h20 UT. The NIR detector was read
%  out after 20 exposures of 11s. The signal-to-noise ratio (S/N)
%is the average measured in each arm. Resolving power $R$ is defined as
%$\lambda / \triangle\lambda = c / \triangle v$.}
%The object was nodded on the 11~arcsec slit
%\label{tab:log}
%\end{center}
%\end{table}

 \section{VLT/X-shooter observations of B275}

VLT/X-shooter spectra were obtained of the massive YSO B275 in M17
\citep[CEN24, RA(2000.0)~$=18^{\rm h}20^{\rm m}25\fs13$, DEC(2000.0)~=~$-16^{\circ}10'24\farcs56$, $V=15.55$~mag, $K=8.05$~mag, ][]{Chini80, Skrutskie06} on August 11,
2009 at 03h20 UT, during the first science verification run (PI Chini). The
observations in the UVB arm (300--600~nm) were binned (2 pixels) in the
wavelength direction in order to increase the signal-to-noise ratio of
this part of the spectrum, while still oversampling the
resolution element. The 1.6$"$ slit was used resulting in resolving
power $R = 3300$. For the VIS (550--1000~nm) and the NIR arm (1000--2500~nm) a
0.9$"$ slit was used ($R = 8800$ and $5600$, respectively). 
The total exposure time was 45 minutes, resulting in a typical signal-to-noise ratio of 70.
For more details on the X-shooter instrument and
its performance, see \citet{D'Odorico06,Vernet11}. The
observing conditions were good (0.6$"$ seeing in $V$ and 76~\%
Moon illumination). The spectra were obtained by nodding the star on
the slit, allowing for background subtraction. The
standard procedures of data reduction were applied using the X-shooter pipeline
version 0.9.4 \citep{Goldoni06,Modigliani10}. For flux calibration and
telluric absorption correction, the standard stars
EG274 and HD180699 were used.

%
%raw science frames were bias-substracted, flatfielded, corrected for
%cosmic ray hits and wavelength calibrated 

\section{Results}

In the following we present
the results for the accurate classification of the photospheric
spectrum, analyze the interstellar spectrum to determine the extinction, 
model the SED using the
flux-calibrated X-shooter spectrum, and describe the emission-line
spectrum produced by the circumstellar disk.

%The optical to near-infrared spectrum covers many hydrogen, helium 
%and metal absorption lines. It also exhibits a number of emission lines: the
%strongest hydrogen lines, the Ca~{\sc ii} triplet lines and the O~{\sc
%  i} line at 845~nm include a double-peaked, central emission feature
%(see Fig.~\ref{fig:spectrum}). CO first- and
%second-overtone emission is present at 1.5 and 2.3~$\mu$m, respectively. The
%interstellar medium is detected in atomic resonance lines and diffuse
%interstellar bands (DIBs). Nebular emission lines are removed during
%the sky-background subtraction procedure. 

%The blue region from 380 to 460~nm, often used to classify early-type
%stars, is shown in Fig.~\ref{fig:spectrum}.
%some forbidden lines are detected (e.g., [Fe~{\sc i}], [O~{\sc i}]), as well as
%(from the Balmer jump at 365~nm, to the Paschen series, up to Brackett $\gamma$ at 2170~nm)
\subsection{Spectral classification}

Hydrogen absorption lines were detected by \citet{Hanson97} in the
blue spectrum of B275, but do not allow for an accurate spectral
classification. As shown in Fig.~\ref{fig:spectrum}, a number of
helium and metal lines can be used to classify the photospheric
spectrum. The He~{\sc i} 400.9~nm and C~{\sc ii} 426.7~nm, prominent
down to spectral type B3, are very weak. The He~{\sc i} 447.1~nm /
Mg~{\sc ii} 448.1~nm ratio is a useful spectral indicator for mid- to
late-B stars \citep{Gray09} as the neutral helium line disappears
towards lower temperature (A0) and the magnesium line
strengthens. When also considering another line ratio, Si~{\sc ii}
412.8~nm / He~{\sc i} 448.1~nm, the spectral type becomes
B6 ($\pm$ one subtype). 

\begin{figure}
   \centering
\includegraphics[width=9cm]{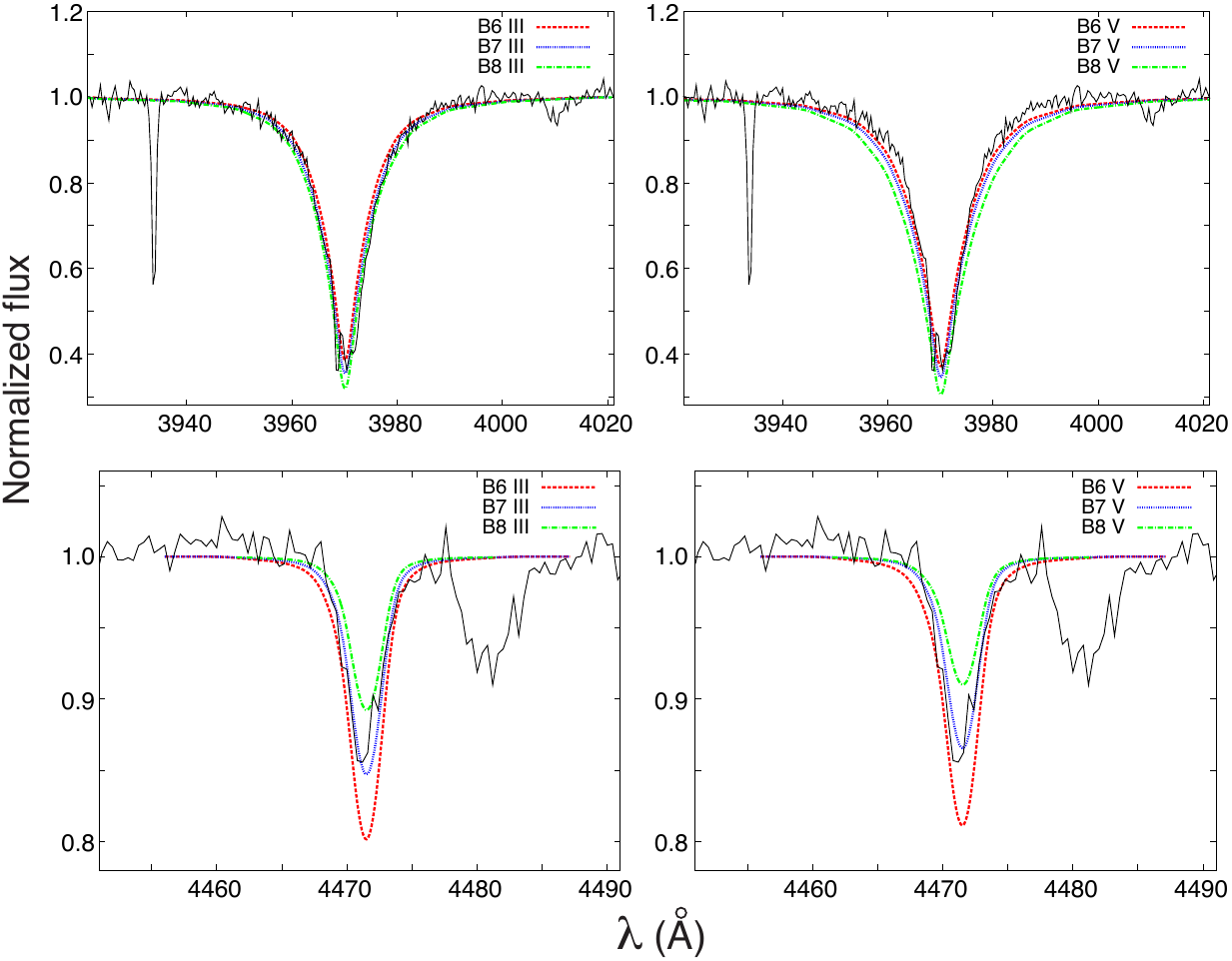}   
\caption{FASTWIND model profiles of He~{\sc i} 447.1~nm ({\it left}) and
     H$\epsilon$ ({\it right}) lines for B6--B8 main-sequence stars ({\it top}) and
     giants ({\it bottom}). The B7~III model provides the best fit with the observed
     profiles.}
              \label{fig:fastwind}
   \end{figure}

The spectral type and luminosity class of B275 are further constrained
by comparison of the observed H~{\sc i} and He~{\sc i} line
profiles (as well as the shape of the SED, see \S~3.3), to model
profiles produced with FASTWIND \citep{Puls05}. This code calculates
non-LTE line-blanketed stellar atmosphere models and is especially
suited to modeling stars with strong winds, but it can also be used to
examine $T_{\rm eff}$ and $\log{g}$ dependent photospheric lines of H
and He.  We constructed a grid of models (in varying $T_{\rm eff}$ and
$\log{g}$) of B6--B8 dwarf and giant stars. The synthetic H~{\sc
  i} and He~{\sc i} profiles resulting from the models are convolved
with the corresponding instrumental and rotational profiles. We adopt
$v_{\rm r} \sin{i} = 100$~km~s$^{-1}$. An acceptable fit is obtained for a B7~V model
(Fig.~\ref{fig:fastwind}); however, the best fit is obtained for a
B7~III model, with $T_{\rm eff}=13,000 \pm 500$~K and $\log{g} = 3.5 \pm
0.3$. This is the first time that the spectral type of a candidate massive YSO has been 
accurately determined.

%Obviously, one would not expect an evolved B star to be member
%of a young star forming region; the importance of the luminosity class
%determination becomes clear in \S~3.3. 

%A proper treatment of the errors is missing here, these error
%estimates are just 1/2 the difference in parameter value between the
%next model in the grid. I do not know how well the others models fit
%%the data.

%, corresponding to a 6.5~M$_{\odot}$ main-sequence star
%  with radius $R \simeq 3.5$~R$_{\odot}$ and effective temperature
%  T$_{\rm eff} \simeq 15,000$~K (ref).

%Information on the luminosity class can be obtained from the spectral
%line broadening. To exclude helium anomalies, one has to check whether
%the helium absorption line strength corresponds to that of the Balmer
%lines. Down to spectral type B5, the ratio of the Si~{\sc iii} 4522~\AA\
%to He~{\sc i} 4387~\AA\ lines is the best indicator, with the Si~{\sc
%  iii} line strengthening and the He~{\sc i} line declining in
%strength with increasing luminosity; however, the Si~{\sc iii} line is
%not detected {\bf Should we not use an upper limit in EW for the Si
%  line?}. Modeling the wings of the H$\gamma$, H$\delta$, and
%H$\epsilon$ lines shows that a giant spectrum provides a slightly
%better fit than a main-sequence star spectrum (Fig.~XXX). The highest
%detectable Balmer line provides information on the (electron) density
%in the photosphere (Inglis-Teller formula) and results in ?. Balmer
%jump?

%Any evidence for veiling?

\subsection{Interstellar spectrum}

The optical spectrum of B275 includes several interstellar features:
atomic resonance transitions (e.g., Ca~{\sc ii} H\&K, Na~{\sc i} D)
and diffuse interstellar bands (DIBs). The DIB strength provides a
measure of the interstellar extinction. For the DIBs centered at
578.0, 579.7, and 661.4~nm, we measure an equivalent width of 0.063,
0.014, and 0.021~nm, respectively, with a typical error of 10~\%. Using
the relations from \citet{Cox05}, we arrive at an E(B-V) of $1.0 \pm
0.1$~mag. For an average value of $R_{V} = 3.1$, these DIB strengths
yield $A_{V} \simeq 3$~mag of visual extinction. This is less than the
determination of $A_{V} \simeq 6.1$~mag from dereddening the SED
(\S~3.3). \citet{Hanson97} note that the DIB features in spectra of
M17 stars do not show large variations in strength, despite the fairly
wide range in total extinction, from $A_{V} = 3 - 10$~mag. We consider
their explanation likely that the DIBs are mostly tracing the
foreground dust and that the (unidentified) DIB carriers may only
exist in the diffuse medium, not in the dark cloud environment of
M17.

   \begin{figure}
   \centering
   \includegraphics[width=8.75cm]{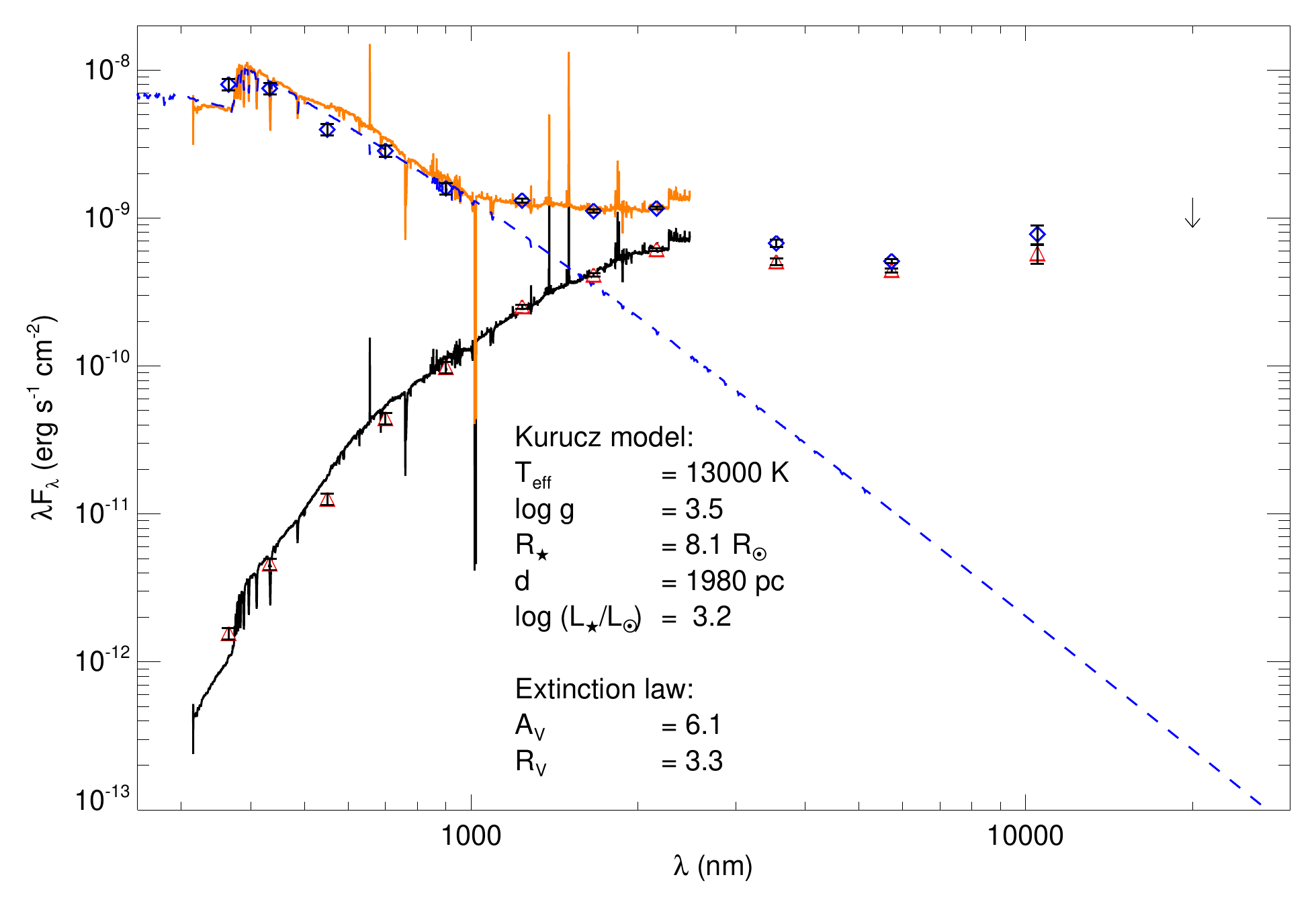}
   \caption{The flux-calibrated X-shooter spectrum of B275 from
     300--2500~nm (black) along with the photometric data (red
     triangles, black error bars) from \citet{Chini80} (UVBRI), 2MASS
     \citep[JHK]{Skrutskie06}, {\it Spitzer} GLIMPSE \citep[3.6, and 5.8~$\mu$m]{Benjamin03}, 
     and \citet{Nielbock01} (N, Q). When dereddened ($A_{V} = 6.1$~mag, orange
     line, blue diamonds), the SED is described well by a
     B7~III Kurucz model (blue, dashed line). The excess flux at 500 -- 800 nm is an instrumental feature.}
     
     %Also the Balmer jump is well
     %fitted. Note the near-infrared excess beyond 1~$\mu$m.}
              \label{fig:SED}
   \end{figure}

\subsection{Spectral energy distribution}

Figure~\ref{fig:SED} shows the flux-calibrated X-shooter spectrum
(300--2500~nm) of B275. The photometric data points demonstrate the accuracy of the spectrophotometric
calibration. The long standing debate over the distance to M17 (ranging from 1.3 to
2.1~kpc) has recently been settled by the measurement of the
trigonometric parallax of the CH$_{3}$OH maser source G15.03--0.68
\citep{Xu11}, resulting in a distance of $1.98^{+0.14}_{-0.12}$~kpc so
that M17 is likely located in the Carina-Sagittarius spiral arm. 

%\citet{Hanson97}, adopting $d = 1.3$~kpc, obtained good correspondence
%with a B0~V Kurucz model for $R_{V} = 3.5$ and $A_{V} = 7-8$~mag. At a
%distance of 2.1~kpc \citep{Hoffmeister08} the corresponding spectral
%type would be O8~V. Such an early spectral type is clearly in
%disagreement with the B6--B7 spectral type derived from the
%photospheric spectrum. 
%Following a similar procedure, 

We deredden the flux-calibrated X-shooter spectrum of B275 (Fig.~\ref{fig:SED})
using the parameterization of the extinction law by
\citet{Cardelli89}. The dereddened spectrum is fit to a Kurucz model \citep{Kurucz79,Kurucz93} based on an iterative 
procedure, with fixed parameters $T_{\rm eff} = 13,000$~K, $\log g = 3.5$,
$d=1.98$~kpc and $R_{V} = 3.3$ (an effective value resulting from interstellar and local extinction). This yields independent best-fit values of $A_{V} = 6.1  \pm 0.6$~mag and $R_\star = 8.1\pm 0.8 $~$R_{\odot}$.  Note that this radius is much larger than that of,
e.g., a B5 zero-age main sequence (ZAMS) star \citep[$2.7 $~$R_{\odot}$, ][]{Hanson97}.  An
additional constraint is provided by the height of the Balmer jump,
which also varies with $T_{\rm eff}$ and $R_\star$: Fig.~\ref{fig:SED}
demonstrates that the Balmer jump (as well as the Paschen jump) is
nicely fit to the observed spectrum. Thus, with a larger radius the
discrepancy between the classification of the photospheric spectrum
and the dereddened SED is solved. The consequence is that B275 is not
on the main sequence but is a so-called bloated star, where the appropriate spectral type would be
B7~III. The corresponding luminosity is $\log{L_\star/L_{\odot}} = 3.2$. 

%The position of B275 in the Hertzsprung-Russell diagram is shown in
%Fig.~\ref{fig:HRD}.

%In this process, the
%stellar radius $R_\star$ is a free parameter; the best fit is obtained for $R_\star
%= 8.1\pm 0.8$~R$_{\odot}$.

% The estimated errors on these values are $\sim$10\% due to the 
% uncertainty of the flux calibration. 

%, which is consistent with $\log{g} = 3.5$ from the line fits (\S~3.1)

%What
%is the radius of luminosity class IV and III?  B5 V 5.9 Msun 3.9 Rsun
%Schmidt-Kaler 1982 Conti 5 Msun B5 III 7 Msun 8 Rsun

%Dereddening of the SED is relatively straightforward in
%the case of massive stars, as the optical and near-infrared part
%correspond to the Rayleigh-Jeans tail of the energy distribution. 

\subsection{Accretion signatures}

%   \begin{figure}
%   \centering
%   \includegraphics[width=8cm]{spectralfeatures.jpg}
%   \caption{The strongest emission lines detected in B275 on a
%     velocity scale. The peak separation varies from $71 \pm
%     7$~km~s$^{-1}$ (O~{\sc ii} 844.6~nm) to $105 \pm 3$~km~s$^{-1}$
%     (Ca~{\sc ii} 849.8~nm). }
%              \label{fig:emlines}
%   \end{figure}

A pronounced, double-peaked emission feature is detected in the
strongest H Balmer lines, the Ca~{\sc ii} triplet and the O~{\sc i}
844.6~nm line (Fig.~\ref{fig:spectrum}). The measured peak-to-peak
separation ranges from $71 \pm 7$~km~s$^{-1}$ (O~{\sc ii} 844.6~nm) to
$105 \pm 3$~km~s$^{-1}$ (Ca~{\sc ii} 849.8~nm), and is centered at the
rest-frame velocity of the star. The Ca~{\sc ii} triplet lines are
probably produced in an optically thick medium, since their strength ratio
is not 1:9:5. The strongest lines of the H~{\sc i} Paschen
and Brackett series also exhibit a central emission component, though it is
single-peaked. The higher series members include a weaker emission
component that may be double-peaked. A number of metallic emission 
lines (e.g., C~{\sc i} and Fe~{\sc ii}) are 
detected throughout the spectrum, albeit very weak.

Prominent CO 1$^{\rm
  st}$-overtone emission bandheads are detected at 2.3~$\mu$m, with
clear evidence of a blue shoulder. We also confirm the presence of
2$^{\rm nd}$-overtone CO bandhead emission at 1.5~$\mu$m
\citep{Hanson97}. CO is easily dissociated so must be shielded from
the strong UV flux of the young massive star. On the other hand, to produce 1$^{\rm st}$ overtone emission,
CO must be excited, requiring a temperature in the range between 1500
and 4500~K \citep{Bik04}. This temperature might even be higher, 
considering the unprecedented detection of 2$^{\rm nd}$ overtone emission. 
These conditions can be met in the plane of a dense
circumstellar disk where the CO molecules can be formed, excited, and
protected from dissociation through self-shielding. The relative
strength and shape of the CO bandheads can be modeled by an optically
thin Keplerian disk \citep{Bik04,Blum04}, where the blue shoulder
would imply a relatively high inclination angle of the disk
(``edge-on''). \citet{Blum04} model the CO 2--0 first-overtone
ro-vibrational bandhead at 2294~nm of B275 resulting in $v\sin{i}
= 109.7 \pm 0.6$~km~s$^{-1}$ (at the inner edge of the CO emission
zone) and surface density $N_{\rm CO} = 3.5 \pm 0.2 \times
10^{21}$~cm$^{-2}$. The double-peaked emission profiles, as shown in
Fig.~\ref{fig:spectrum}, are very similar to the emission-line profile
of a single line obtained by \citet{Blum04}.

We find no evidence for veiling of the optical spectrum or any
strong indications of active ``heavy'' accretion and/or jets, such as those
observed in some other systems \citep[e.g., ][]{Ellerbroek11}. The [O~{\sc ii}]
    630~nm line very likely has a nebular origin. 

   \begin{figure}
   \centering
%   \hbox{
   \includegraphics[width=7.75cm]{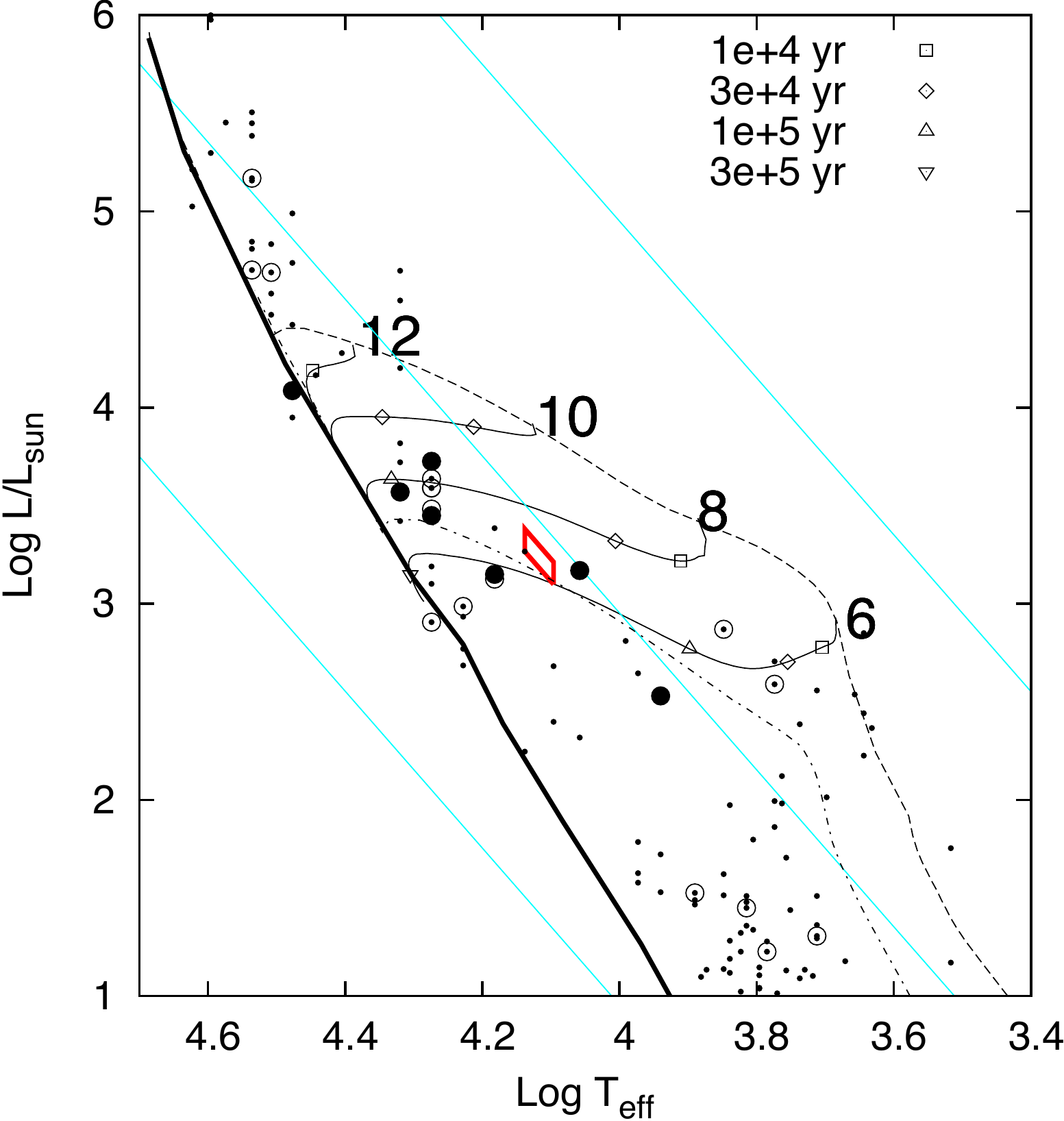}
   \caption{The location of B275 (red parallellogram) in the HRD next
     to PMS tracks from \citet{Hosokawa09} with the ZAMS
     mass labeled and open symbols indicating lifetimes. The thin dashed and thin dot-dashed lines are the birth lines for 
     accretion rates of 10$^{-4}$~M$_{\odot}$~yr$^{-1}$ and 10$^{-5}$~M$_{\odot}$~yr$^{-1}$, respectively; the thick solid
     line is the ZAMS \citep{Schaller92}. The filled and open circles represent stars in
     M17 for which a spectral type has been determined
     \citep{Hoffmeister08}, within a radius of $0 \farcm 5$ and $1
     \farcm 0$, respectively; dots are other stars in M17. B275 is on its way to becoming a
     6--8~M$_{\odot}$ ZAMS star.}
              \label{fig:HRD}
   \end{figure}

\section{Discussion}

The accurate spectral classification and SED fit result in a
well-defined position of B275 in the Hertzsprung-Russell diagram (HRD,
Fig.~\ref{fig:HRD}). It is located well above the ZAMS, demonstrating
its PMS nature. If B275 is contracting
towards the ZAMS, the final ZAMS mass would be
6--8~M$_{\odot}$ (spectral type B1-B2), assuming that no additional mass is accreted.

% Besides the presence of a circumstellar disk, no evidence is found for ongoing accretion. 
% (thick dashed line)

%Fig.~\ref{fig:HRD} includes evolutionary tracks of PMS stars following an accretion rate of
%10$^{-4}$~M$_{\odot}$~yr$^{-1}$ \citep{Hosokawa09}; the thin dashed line represents the
%so-called birthline, the solid lines the PMS tracks after mass
%accretion has ceased \citep{Hosokawa09}.  

To be visible at this location in the HRD, the star must have experienced 
an average accretion rate of at least 10$^{-5}$~M$_{\odot}$~yr$^{-1}$ in its recent history.
B275 may thus be the long-sought-for example of an early-B PMS star. Figure~\ref{fig:HRD} also shows other nearby stars in M17. \citet{Hanson97} derive an age of
$\sim$1~Myr for the M17 cluster and \citet{Hoffmeister08} estimate that the PMS
objects are less than $5 \times 10^{5}$~yr old. Based on its location on the HRD we estimate the age of B275 at  $10^{5}$~yr.

%These objects are hard to find because of the very short
%Kelvin-Helmholtz timescale of $\sim 10^{4}$~yr (??? compare to lifetimes).

%B275 would be such a young star,
%not yet having photo-evaporated its circumstellar disk. 

%B275 is located above the birthline corresponding to an accretion rate
%of 10$^{-5}$~M$_{\odot}$~yr$^{-1}$ \citep{Hosokawa09}. 

B275 bears some resemblance to a classical Be star. However, we note that Be stars do not emit CO $1^{\rm st}$ and $2^{\rm
  nd}$ overtone emission. In addition, B275 would be classified as a luminous Herbig Be star according to the definition discussed in \citet{Carmona}, but our analysis of the photospheric spectrum allows for a more quantitative classification.

%  Be stars would rotate at or near the critical break-up speed; the
%metallic photospheric lines of B275 are not significantly rotationally
%broadened. Considering the inclination implied by the blue shoulder 
%in the CO overtone emission,
%this reveals that the photosphere is not rotating at the typical velocity of a Be-star. 

%We classify B275 as a massive PMS star, similar to the 
%more massive stars in the Herbig Be sample \citep{The94}.

%Herbig Be stars often appear rather isolated, relatively close by and
%possess very long lived disks \citep{Waters98}. B275 is different,
%because of its high location in the HRD and the fact that it is
%located in a massive star forming cluster.

% and the apparent cease of (or pause in) the accretion process. 

%The
%simultaneous detection of the photosphere and the remnant of a possibly
%massive accretion disk raises questions about the lifetime of
%circumstellar disks around massive YSOs.

B275 has a significant amount of infrared excess, starting at 1~$\mu$m, and 
a flat SED between 2 and 10~$\mu$m \citep{Nielbock01}. This indicates the 
presence of a flaring circumstellar disk in which the dust has not settled yet. 
However, the visibility of the photosphere, the small number of optical gas emission lines 
and the absence of a jet lead us to believe that the current mass-accretion 
rate is not very high. Nevertheless, the CO (2-0) and (3-0) emission 
originates in a dense and highly excited inner part of the disk. Either the system 
is in an intermittent phase between accretion episodes, or it is on the verge 
of photo-evaporating its disk. Either scenario is consistent with its location in the 
HR-diagram: an intermediate-mass, visible star in M17 on its way to becoming an early-B main-sequence star.

\begin{acknowledgements}
  We thank Takashi Hosokawa for kindly providing the PMS tracks. The
  ESO Paranal staff is acknowledged for obtaining the X-shooter
  spectrum of B275. We thank the anonymous referee for useful comments and suggestions.
\end{acknowledgements}

\bibliographystyle{aa}

\bibliography{M17_Xshooter}

\end{document}